\documentclass[conference]{IEEEtran}
\IEEEoverridecommandlockouts

\usepackage{cite}
\usepackage{amsmath,amssymb,amsfonts}
\usepackage{algpseudocode}
\usepackage{graphicx}
\usepackage{textcomp}
\usepackage{xcolor}
\usepackage{algorithm}
\usepackage{booktabs}
\usepackage{multirow}
\usepackage{bm}
\def\BibTeX{{\rm B\kern-.05em{\sc i\kern-.025em b}\kern-.08em
    T\kern-.1667em\lower.7ex\hbox{E}\kern-.125emX}}

\begin{document}

\title{SkySense: A Semi-Supervised Generative Framework for UAV Localization in ISAC Networks\\
}

\author{%
\IEEEauthorblockN{%
 Shenghan Luo$^1$, Yin Xu$^2$, Jie Yang$^3$, Yang Wang$^2$, Cixiao Zhang$^2$, Shi Jin$^1$, Wenjun Zhang$^2$ %
}
\IEEEauthorblockA{%
$^1$ School of Information Science and Engineering, Southeast University, Nanjing 211189, China\\
$^2$ Cooperative Medianet Innovation Center (CMIC), Shanghai Jiao Tong University, Shanghai 200240, China\\
$^3$ School of Automation, Southeast University, Nanjing 210096, China\\
Corresponding author: Yin Xu (e-mail: xuyin@sjtu.edu.cn)%
}
}

\maketitle

\begin{abstract}
Extreme data scarcity and inherent multipath spatial ambiguity severely limit existing deep learning-based channel state information (CSI) fingerprinting localization schemes for target unmanned aerial vehicles (UAVs). To overcome these challenges, we propose an end-to-end semi-supervised generative localization framework. First, by exploiting the temporal correlations inherent in continuous flight trajectories, a self-supervised encoder extracts robust spatial features from massive unlabeled CSI sequences to establish structured latent representations.  Following this, we utilize a consistency model, a powerful derivative of diffusion architectures, as the core generative backbone to map the learned latent space to physical coordinates, jointly fine-tuning the pre-trained encoder with a strictly limited set of labeled CSI. This consistency formulation models the conditional distribution to resolve the mean collapse problem of discriminative models, while compressing the inference trajectory to 1-2 steps to avoid the latency bottleneck of traditional diffusion models. Furthermore, a lightweight distributed fusion mechanism is designed to aggregate spatial predictions across multiple base stations (BS) from a multi-view geometry perspective. Comprehensive evaluations on a real-world measurement dataset demonstrate that our framework achieves low latency and suppresses the mean localization error to 9.77 cm under a 3-BS fusion setup with only a 1\% label fraction, significantly outperforming existing fully supervised and semi-supervised discriminative baselines.
\end{abstract}


\section{Introduction}

With the rapid proliferation of the low-altitude economy, the ubiquitous deployment of unmanned aerial vehicles (UAVs) has raised critical concerns regarding airspace security, making the precise localization of unauthorized UAVs an urgent necessity \cite{11259061, 11131292}. In the emerging sixth-generation (6G) integrated sensing and communication (ISAC) paradigm \cite{11098638}, utilizing existing communication infrastructure for the passive sensing of such non-cooperative targets has become a highly promising solution. By exploiting channel state information (CSI) as unique spatial fingerprints, passive localization effectively captures the target's unique scattering signature embedded within the rich multi-path environment. Consequently, CSI-based fingerprinting is regarded as a prominent technology to satisfy the demanding requirements of 6G ISAC networks \cite{11456648}.

Despite its great potential, practical UAV localization using CSI is fundamentally challenged by spatial ambiguity, where distinct physical positions can induce highly similar CSI fingerprints. When confronted with such overlapping mappings, conventional discriminative neural networks tend to yield invalid average coordinates, resulting in severe performance degradation. Unlike discriminative methods, generative diffusion models have been introduced as a highly promising solution by probabilistically modeling the conditional distribution to reliably resolve spatial ambiguities \cite{10599123}. However, their reliance on a long-chain Markov iterative denoising process incurs prohibitively high inference latency, making them computationally infeasible for highly dynamic UAV tracking. While recent advanced generative techniques, specifically consistency models \cite{song2023consistency, song2024improved}, offer a paradigm shift to bypass this lengthy iterative process, how to effectively exploit their generative capability for real-time UAV localization remains an open challenge.

Beyond the inherent spatial ambiguity and inference latency, the real-world deployment of these advanced deep learning architectures is severely hindered by an extreme data scalability bottleneck. Previous foundational works utilizing multi-layer perceptrons (MLPs) \cite{7438932}, convolutional neural networks (CNNs) \cite{8027020}, and most recently, FC-AE-GPR \cite{10636967}, typically adopt fully supervised learning paradigms, which heavily rely on constructing massive fingerprinting databases strictly paired with precise centimeter-level real-time kinematic (RTK) coordinates \cite{zhu2025csibench}. The acquisition of such physical annotations is prohibitively expensive, especially for 3D localization tasks, thereby creating a severe barrier for practical ISAC deployments. To alleviate this dependency, semi-supervised learning paradigms that leverage massive, effortlessly acquired unlabeled data have emerged as a prominent direction, represented by recent state-of-the-art architectures such as SSLUL \cite{10555943}. However, these conventional semi-supervised approaches extract features primarily by minimizing CSI reconstruction errors, a generic data-recovery objective that inherently lacks explicit guidance for spatial mapping.

To simultaneously address the spatial ambiguity of discriminative models and the extreme data hunger of practical deployments, we propose SkySense, a semi-supervised consistency generative framework for 3D UAV localization. Motivated by the physical principle that temporally adjacent UAV observations are inherently proximate in space, we design a self-supervised training strategy. This strategy compels the encoder to extract robust spatial representations from massive unlabeled CSI sequences, effectively mitigating the severe label dependency. Furthermore, to overcome the multi-path ambiguity and latency bottlenecks, a conditional consistency model is seamlessly integrated as the core generative backbone. This architecture explicitly models the conditional spatial distribution to reliably avoid mean collapse, while compressing the inference trajectory to merely 1-2 steps. Finally, to bypass the prohibitive overhead of joint multi-BS training, we build upon this robust generative foundation to introduce a lightweight distributed fusion mechanism that aggregates spatial predictions across multiple base stations. Experimental evaluations on an open-source measurement dataset demonstrate that the proposed SkySense delivers exceptional performance, suppressing the 3D localization error to an unprecedented 9.77 cm under a 3-BS fusion setup with only a 1\% label fraction, significantly outperforming existing fully supervised regression and standard diffusion baselines.

\section{System Model}

\begin{figure}[htbp]
    \centering
    \includegraphics[width=0.9\linewidth]{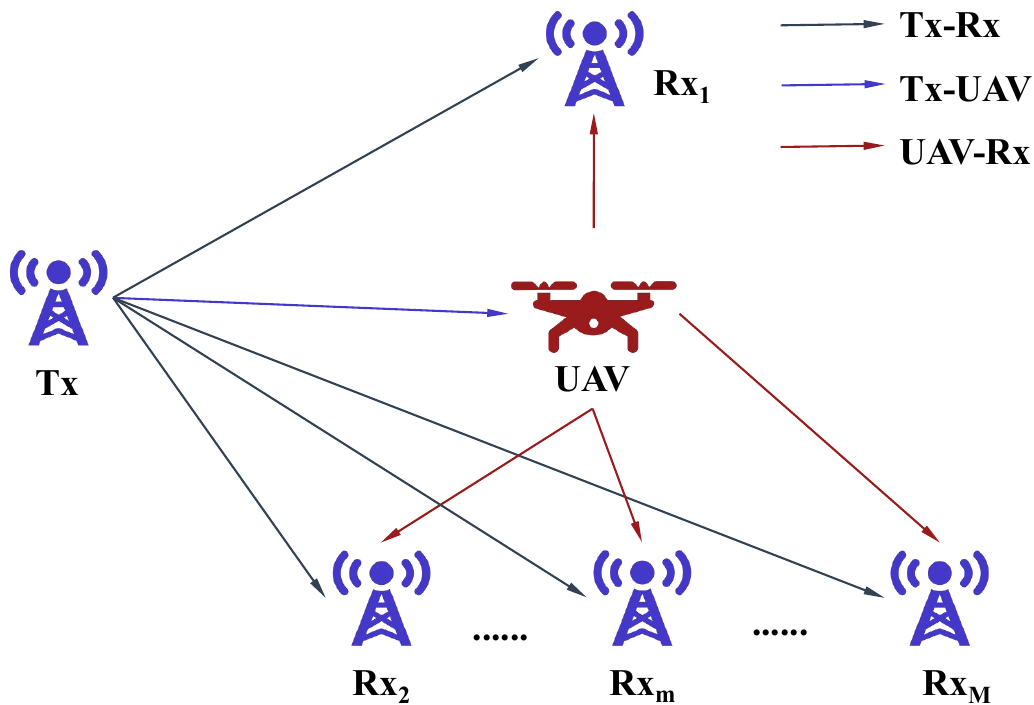}
    \caption{An illustration of the multi-static ISAC network deployed for unauthorized UAV localization.}
    \label{fig:scenario}
\end{figure}

Consider a multi-BS ISAC network, aiming to passively localize an unauthorized and silent UAV, as illustrated in Fig.~\ref{fig:scenario}. In this scenario, the transmitting BS continuously broadcasts ambient communication signals, which simultaneously illuminate the surveillance area and are scattered by the non-cooperative target UAV. A set of $M$ spatially distributed receiving BSs, indexed by $m \in \{1, 2, \dots, M\}$, captures these scattered echoes, continuously recording the CSI induced by the continuous flight of the UAV. We assume that all BSs are physically stationary with precisely surveyed geometric locations. To maintain analytical tractability, we restrict our scenario to a single unauthorized UAV maneuvering dynamically within the region of interest. The unknown 3D position vector of the silent target UAV is represented as $\mathbf{x}_{0}$.

\subsection{Signal Model}
In ISAC Networks, the transmitting base station broadcasts orthogonal frequency-division multiplexing (OFDM) probing signals across $K$ subcarriers. At the $m$-th distributed receiver, the captured frequency-domain signal vector on the $k$-th subcarrier ($k \in \{1, 2, \dots, K\}$) is formulated as:
\begin{equation}
    \mathbf{R}_k^{(m)} = \widetilde{\mathbf{H}}_k^{(m)} \mathbf{x}_k + \mathbf{W}_k^{(m)},
\end{equation}
where $\mathbf{R}_k^{(m)} \in \mathbb{C}^{N_{rx} \times 1}$ denotes the received signal, $\mathbf{x}_k \in \mathbb{C}^{N_{tx} \times 1}$ is the known transmitted OFDM symbol, and $\mathbf{W}_k^{(m)}$ represents the additive white Gaussian noise. The matrix $\widetilde{\mathbf{H}}_k^{(m)} \in \mathbb{C}^{N_{rx} \times N_{tx}}$ characterizes the channel frequency response (CFR) linking the $N_{tx}$ transmitting antennas and $N_{rx}$ receiving antennas. 

Assuming precise synchronization and ideal channel estimation, the receivers can accurately acquire these CFR matrices. This assumption allows us to strictly evaluate the efficacy of the downstream localization algorithms without the compounding effects of estimation errors. Aggregating these matrices along the frequency dimension forms a 3D CFR tensor, denoted by $\widetilde{\boldsymbol{\mathcal{H}}}^{(m)} \in \mathbb{C}^{N_{rx} \times N_{tx} \times K}$.

To extract stable signatures, we map the frequency-domain observations into the delay domain via an Inverse Fast Fourier Transform (IFFT). Let $\mathbf{F}^{-1} \in \mathbb{C}^{K \times K}$ represent the standard unitary IFFT matrix. For antenna pair $(u, v)$, the corresponding channel impulse response (CIR) vector is computed as:
\begin{equation}
    [\boldsymbol{\mathcal{H}}_{\text{CIR}}^{(m)}]_{u,v,\tau} = \sum_{k=1}^{K} [\mathbf{F}^{-1}]_{\tau,k} [\widetilde{\boldsymbol{\mathcal{H}}}^{(m)}]_{u,v,k}, \quad \tau \in \{1,\dots,K\}.
\end{equation}

Recognizing that the dominant multi-path energy is clustered within the earliest arrival paths, we exploit this sparsity by truncating the CIR tensor. By retaining only the first $L$ effective delay taps ($L \ll K$), we obtain a compact, information-dense tensor $\mathbf{H}^{(m)} \in \mathbb{C}^{N_{rx} \times N_{tx} \times L}$:
\begin{equation}
    [\mathbf{H}^{(m)}]_{u,v,\tau} = [\boldsymbol{\mathcal{H}}_{\text{CIR}}^{(m)}]_{u,v,\tau}, \quad 1 \leq \tau \leq L.
\end{equation}

Finally, this complex-valued tensor is decoupled into its amplitude component $\mathbf{A}^{(m)} = |\mathbf{H}^{(m)}|$ and phase component $\mathbf{\Phi}^{(m)} = \angle(\mathbf{H}^{(m)})$. These are vectorized and concatenated as input of localization framework:
\begin{equation}
    \mathbf{h}^{(m)} = \begin{bmatrix} vec(\mathbf{A}^{(m)}) \\ vec(\mathbf{\Phi}^{(m)}) \end{bmatrix} \in \mathbb{R}^d,
\end{equation}
where $d = 2 N_{rx} N_{tx} L$.

\subsection{Problem Statement}
The data acquired at the $m$-th base station is partitioned into a highly limited labeled subset $\mathcal{D}^{(m)}_L = \{(\mathbf{h}^{(m)}_i, \mathbf{x}_{i,0})\}_{i=1}^{N_L}$ and a massive unlabeled sequential subset $\mathcal{D}^{(m)}_U = \{\mathbf{h}^{(m)}_i\}_{i=1}^{N_U}$, where $N_L \ll N_U$. Unlike traditional discriminative models that optimize a point estimate $\hat{\mathbf{x}} = f_\omega(\mathbf{h}^{(m)})$, the learning objective of generative localization paradigm is to accurately model the conditional probability distribution $p(\mathbf{x}|\mathbf{h}^{(m)})$. 

\section{End-to-end Localization Framework}
\label{sec:framework}
\begin{figure*}[h]
    \centering
    \includegraphics[width=0.92\textwidth]{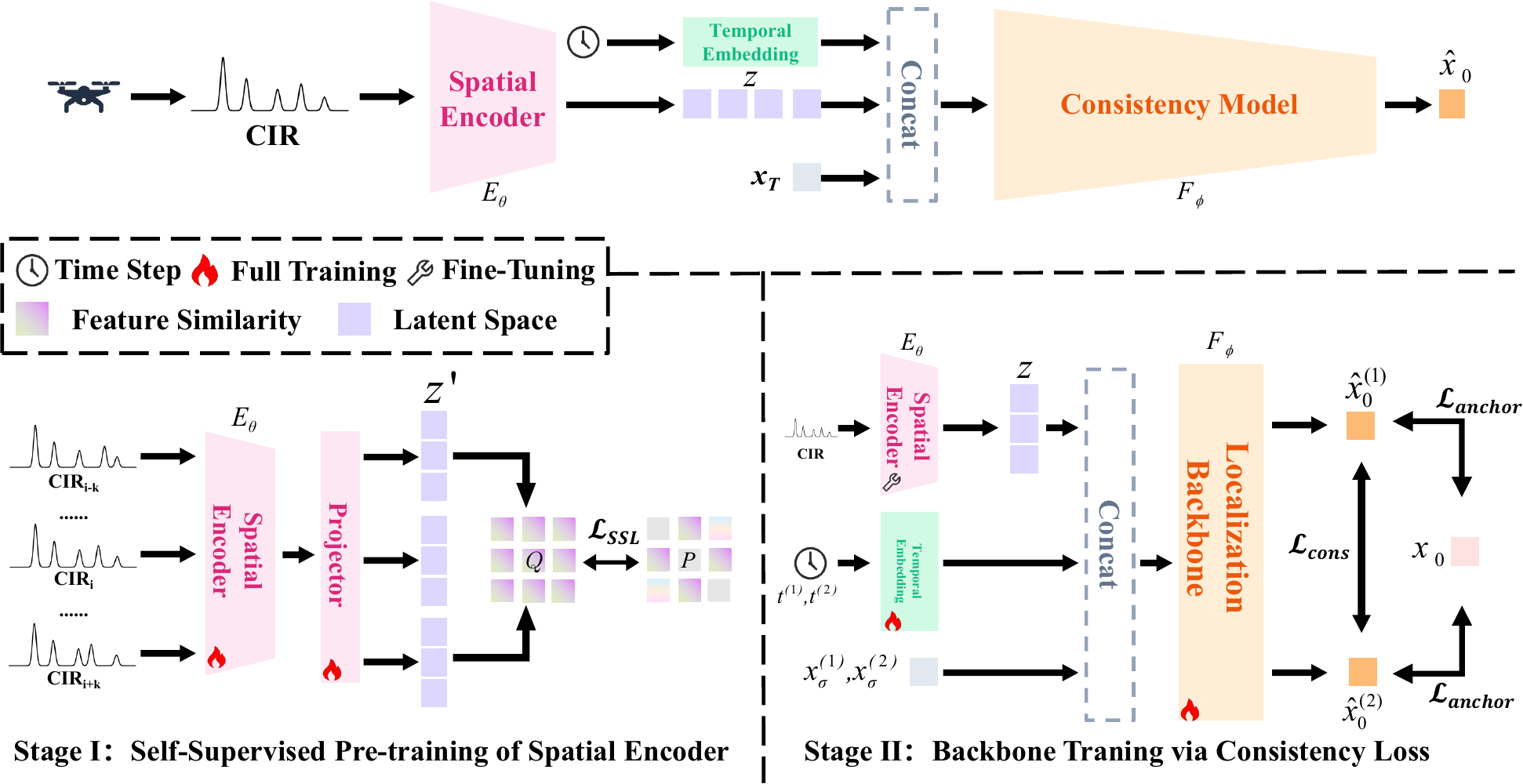}
    \caption{The overall architecture of the proposed SkySense framework. The learning paradigm consists of two stages: Stage I extracts the underlying spatial manifold representations from continuous unlabeled CSI sequences via self-supervised temporal contrastive learning; Stage II maps these compact spatial features to exact physical coordinates utilizing a conditional consistency generative backbone.}
    \label{fig:framework}
\end{figure*}
This section details the proposed SkySense framework. As illustrated in Fig.~\ref{fig:framework}, SkySense comprises a two-stage process: Stage I extracts spatial representations via self-supervised learning, and Stage II maps these representations to physical coordinates utilizing a conditional consistency generative backbone. A key design feature of SkySense is that it performs offline training locally on single-BS measurements, while fully supporting cooperative multi-BS fusion during online inference. Accordingly, the node index superscript $(m)$ is omitted for notational simplicity when formulating the local training objectives in Sections~\ref{subsec:temporal_pretraining} and \ref{subsec:consistency_finetuning}, and is reintroduced in Section~\ref{subsec:online_inference} to describe the distributed fusion process.

\subsection{Self-Supervised Spatial Representation Extractor}
\label{subsec:temporal_pretraining}
Motivated by the kinematic continuity of the UAV's flight trajectory, observations sampled at close temporal intervals exhibit strong spatial correlations. We exploit these correlations to learn robust spatial representations from massive unlabeled CSI sequences $\mathcal{D}_U$ by employing a neural encoder $E_\theta$ to extract compact features $\mathbf{z}_i = E_\theta(\mathbf{h}_i)$ in the latent space.

Subsequently, a non-linear projector network maps these representations into an auxiliary projection space to produce $\mathbf{z}'_i$. During the pre-training phase, we construct a spatial similarity prediction matrix $\mathbf{Q}$, where its element $Q(j|i)$ represents the predicted spatial similarity between the $i$-th and $j$-th frames, obtained via a softmax function:
\begin{equation}
    Q(j|i) = \frac{\exp\left( \frac{{\mathbf{z}'_i}^T \mathbf{z}'_j}{\tau \|\mathbf{z}'_i\|_2 \|\mathbf{z}'_j\|_2} \right)}{\sum_{k \neq i} \exp\left( \frac{{\mathbf{z}'_i}^T \mathbf{z}'_k}{\tau \|\mathbf{z}'_i\|_2 \|\mathbf{z}'_k\|_2} \right)},
\end{equation}
where $\tau$ is a temperature hyperparameter. Concurrently, we leverage the timestamps of the continuous sequences to construct a temporal target matrix $\mathbf{P}$. To reflect the property that temporally closer frames share higher spatial similarity, its element $P(j|i)$ serves as the temporal prior and is modeled via a Gaussian decay function:
\begin{equation}
    P(j|i) = \frac{w_{i,j}}{\sum_{k \neq i} w_{i,k}}, \quad w_{i,j} = \exp\left(-\frac{(i-j)^2}{2\rho_{t}^2}\right),
\end{equation}
where $\rho_{t}$ controls the temporal receptive field and the diagonal elements are masked ($w_{i,i} = 0$) to prevent trivial self-matching. 

Finally, we align the network's spatial similarity predictions with the temporal priors by minimizing the expected soft-target Cross-Entropy over the data distribution:
\begin{equation}
    \mathcal{L}_{\text{SSL}} = -\mathbb{E} \left[ \sum_{j \neq i} P(j|i) \log Q(j|i) \right].
\end{equation}

\subsection{Conditional Consistency Generative Backbone}
\label{subsec:consistency_finetuning}
To overcome the precision collapse and latency bottlenecks of traditional generative models under label scarcity, we formulate the coordinate mapping as a conditional consistency generative process. By learning a direct mapping that transports arbitrary noisy states along a probability flow trajectory to the clean physical origin, our backbone $F_\phi$ utilizes the pre-trained spatial representation $\mathbf{z}_i$ from a limited labeled anchor set $\mathcal{D}_L = \{(\mathbf{h}_i, \mathbf{x}_{i,0})\}_{i=1}^{N_L}$ to achieve high-precision 3D localization, as is summarized in Algorithm \ref{alg:consistency_train}.

Unlike standard diffusion models operating on discrete noise schedules, our consistency framework parameterizes perturbation using a continuous noise scale $\sigma \in [\sigma_{\min}, \sigma_{\max}]$. During the forward pass, for the same clean target $\mathbf{x}_{0}$, we independently sample two continuous noise levels $\sigma^{(1)}, \sigma^{(2)} \sim \mathrm{Uniform}[\sigma_{\min}, \sigma_{\max}]$. These continuous scales are bijectively mapped to normalized time-steps $t^{(i)} = \ln(\sigma^{(i)}/\sigma_{\max}) / \ln(\sigma_{\min}/\sigma_{\max})$ for $i \in \{1, 2\}$ to condition the neural network. Consequently, the corresponding noisy states are generated by injecting scaled Gaussian noise: $\mathbf{x}_{\sigma}^{(i)} = \mathbf{x}_{0} + \sigma^{(i)}\boldsymbol{\epsilon}^{(i)}$, where $\boldsymbol{\epsilon}^{(i)} \sim \mathcal{N}(\mathbf{0}, \mathbf{I})$.

\begin{algorithm}[h]
\caption{Localization Backbone Training via Consistency Loss}
\label{alg:consistency_train}
\begin{algorithmic}[1]
\Require Noise bounds $\sigma_{\min}, \sigma_{\max}$; labeled dataset $\mathcal{D}_L = \{(\mathbf{h}_j, \mathbf{x}_{j,0})\}_{j=1}^{N_L}$; pre-trained encoder $E_\theta$; mini-batch size $B$; learning rates $\eta$ and $\eta_{enc}$ ($\eta_{enc} \ll \eta$)
\State Initialize generative backbone parameters $\phi$
\Repeat
    \State Sample a mini-batch $\{(\mathbf{h}_j, \mathbf{x}_{j,0})\}_{j=1}^{M}$ from $\mathcal{D}_L$
    \For{$j = 1$ to $M$}
        \State $\mathbf{z}_j \gets E_\theta(\mathbf{h}_j)$ \Comment{Extract latent spatial condition}
        \State $\sigma^{(1)}, \sigma^{(2)} \sim \mathrm{Uniform}[\sigma_{\min}, \sigma_{\max}]$
        \For{$i = 1$ to $2$}
            \State $t^{(i)} \gets \dfrac{\ln(\sigma^{(i)}/\sigma_{\max})}{\ln(\sigma_{\min}/\sigma_{\max})}$
            \State $\mathbf{x}_{\sigma^{(i)},j} \gets \mathbf{x}_{j,0} + \sigma^{(i)} \boldsymbol{\epsilon}^{(i)}_j, \boldsymbol{\epsilon}^{(i)}_j \sim \mathcal{N}(\mathbf{0}, \mathbf{I})$
        \EndFor
        \State $\hat{\mathbf{x}}^{(1)}_{j} \gets F_{\phi}(\mathbf{x}_{\sigma^{(1)},j},\, t^{(1)},\, \mathbf{z}_{j})$
        \State $\hat{\mathbf{x}}^{(2)}_{j} \gets F_{\phi}(\mathbf{x}_{\sigma^{(2)},j},\, t^{(2)},\, \mathbf{z}_{j})$
    \EndFor
    \State $\displaystyle \mathcal{L}_{\text{total}} \gets \frac{1}{B}\sum_{j=1}^{B}\Big(
        \|\hat{\mathbf{x}}^{(1)}_{j} - \hat{\mathbf{x}}^{(2)}_{j}\|^2
        + \|\hat{\mathbf{x}}^{(1)}_{j} - \mathbf{x}_{j,0}\|^2
        + \|\hat{\mathbf{x}}^{(2)}_{j} - \mathbf{x}_{j,0}\|^2 \Big)$
    \State $\phi \gets \phi - \eta \nabla_{\phi} \mathcal{L}_{\text{total}}$ \Comment{Update generative backbone}
    \State $\theta \gets \theta - \eta_{\mathrm{enc}} \nabla_{\theta} \mathcal{L}_{\text{total}}$ \Comment{Fine-tune encoder}
\Until{converged on $\mathcal{D}_L$}
\State \Return $\{F_\phi(\cdot,\cdot,\cdot), E_\theta(\cdot)\}$
\end{algorithmic}
\end{algorithm}

Let $F_\phi(\mathbf{x}_\sigma, t, \mathbf{z})$ denote our trainable generative backbone, and let $\hat{\mathbf{x}}^{(i)} = F_\phi(\mathbf{x}_{\sigma}^{(i)}, t^{(i)}, \mathbf{z})$ represent the generative spatial prediction from the $i$-th sampled noise state. To strictly align with our decoupled architectural design, the overall optimization objective $\mathcal{L}_{\text{total}}$ is explicitly partitioned into a structural consistency term $\mathcal{L}_{\text{cons}}$ and a physical anchor term $\mathcal{L}_{\text{anchor}}$, formulated as:
\begin{equation}
    \mathcal{L}_{\text{total}} = \mathcal{L}_{\text{cons}} + \mathcal{L}_{\text{anchor}}.
\end{equation}
The structural consistency term is defined as:
\begin{equation}
    \mathcal{L}_{\text{cons}} = \mathbb{E} \left[ \left\| \hat{\mathbf{x}}^{(1)} - \hat{\mathbf{x}}^{(2)} \right\|_2^2 \right].
\end{equation}
This formulation rigorously enforces the self-consistency property, ensuring that predictions originating from distinct noise states along the same ordinary differential equation trajectory map to the identical clean coordinate. Concurrently, the physical anchor term is formulated as:
\begin{equation}
    \mathcal{L}_{\text{anchor}} = \mathbb{E} \left[ \sum_{i=1}^{2} \left\| \hat{\mathbf{x}}^{(i)} - \mathbf{x}_{0} \right\|_2^2 \right].
\end{equation}
This term explicitly anchors these generative predictions to the scarce physical ground truths. During this joint optimization, $F_\phi$ is updated using a standard learning rate $\eta$, while $E_\theta$ is fine-tuned with a significantly reduced learning rate $\eta_{enc}$.

\subsection{Online Inference and Multi-BS Cooperative Fusion}
\label{subsec:online_inference}
Building upon the local single-BS training paradigm, the framework seamlessly extends to support both independent single-BS localization ($M=1$) and multi-BS cooperative fusion ($M>1$) during online inference. First, the frozen encoder extracts spatial representations $\mathbf{z}^{(m)}$ from the $M$ participating base stations. To discretize the continuous probability flow for inference, we define a highly compressed schedule of $S$ steps with strictly decreasing noise levels $\{\sigma_i\}_{i=1}^S \in [\sigma_{\min}, \sigma_{\max}]$ determined by the standard Karras scheduling \cite{song2023consistency, song2024improved} and corresponding timesteps $t_i = \ln(\sigma_i/\sigma_{\max}) / \ln(\sigma_{\min}/\sigma_{\max})$.

The generative process initiates from a shared pure noise state $\mathbf{x}_{\sigma_1} \sim \mathcal{N}(\mathbf{0}, \sigma_{\max}^2 \mathbf{I})$. At each step $i$, every base station $m$ independently utilizes the generative backbone to map the current noisy state directly to a clean coordinate proposal: $\hat{\mathbf{x}}_{m} = F_\phi(\mathbf{x}_{\sigma_i}, t_i, \mathbf{z}^{(m)})$. While a single-BS deployment directly utilizes its local proposal, a multi-BS setup aggregates these distributed predictions via spatial consensus to form a unified origin estimate: $\hat{\mathbf{x}}_{\text{fused}} = \frac{1}{M} \sum_{m=1}^{M} \hat{\mathbf{x}}_{m}$. If further refinement is required ($i < S$), this fused estimate is re-perturbed to the subsequent noise level $\sigma_{i+1}$ to construct the next input state: $\mathbf{x}_{\sigma_{i+1}} = \hat{\mathbf{x}}_{\text{fused}} + \sigma_{i+1} \boldsymbol{\epsilon}$, where $\boldsymbol{\epsilon} \sim \mathcal{N}(\mathbf{0}, \mathbf{I})$. After executing this loop, the final consensus state $\hat{\mathbf{x}}_{\text{fused}}$ serves as the precise 3D position estimate.
\section{Evaluation}
\label{sec:evaluation}

The evaluation utilizes the open-source ISAC UAV Dataset\cite{10133118}, which contains 3-BS radar surveillance measurements of a non-cooperative UAV in complex urban environments. The data sequences are partitioned into disjoint sets using a time-truncated strategy, where the middle 80\% of the continuous flight trajectory serve as the training set, and the initial 10\% and final 10\% segments serve as the testing set. All training and inference procedures are executed on a NVIDIA GeForce RTX 4090 GPU to ensure a consistent computational environment.

\subsection{Overall Localization Performance}
\label{subsec:overall_performance}
For the overall performance evaluation, the proposed SkySense is benchmarked against two representative discriminative baselines, all under a single-BS setting: FC-AE-GPR\cite{10636967} for fully-supervised models and SSLUL\cite{10555943} for semi-supervised models. Within the training phase, label density is controlled by evaluating varying levels of label sparsity, reserving 0.1\% to 1.0\% of the high-fidelity RTK coordinates as labeled anchors. The remaining 99.0\% to 99.9\% of the CSI sequences are utilized exclusively as unlabeled data for the self-supervised pre-training stage. 

\begin{figure}[htbp]
    \centering
    \includegraphics[width=0.8\linewidth]{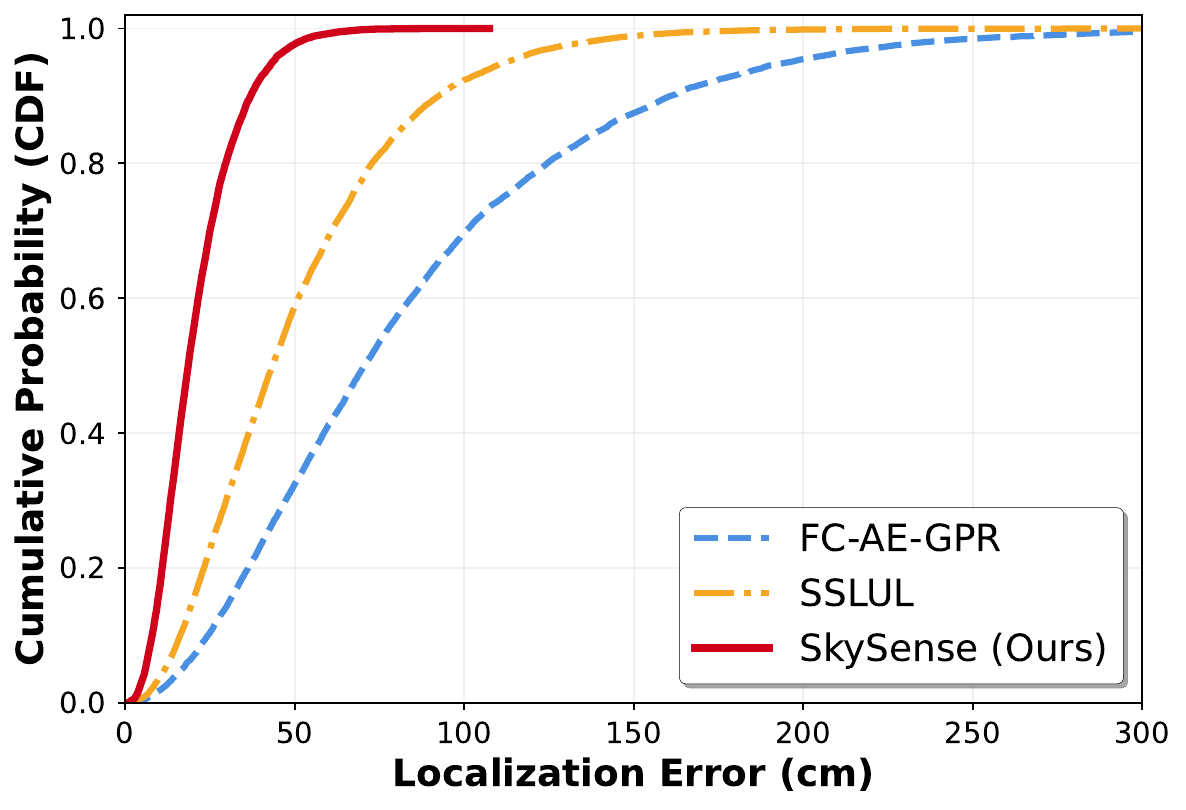}
    \caption{CDF of 3D localization errors under the 1\% label density setting.}
    \label{fig:CDF_Error_1_Percent}
\end{figure}

\begin{figure}[htbp]
    \centering
    \includegraphics[width=0.8\linewidth]{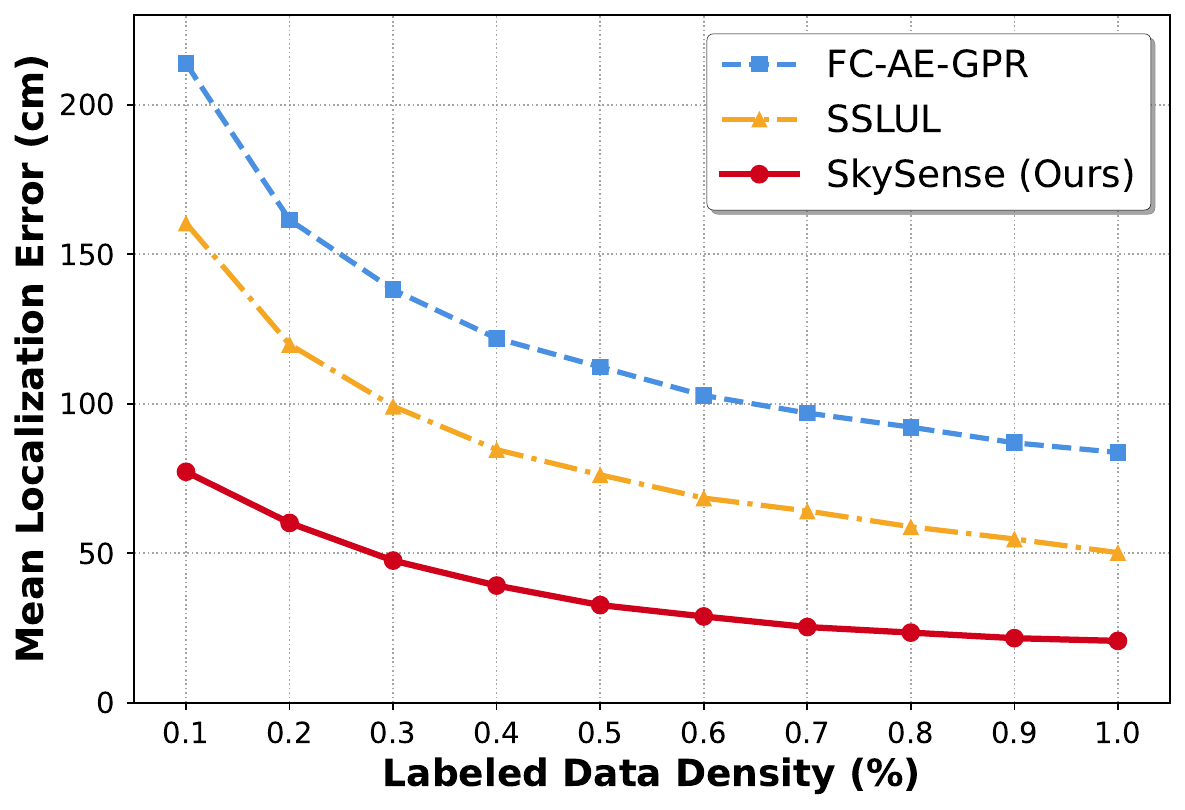}
    \caption{Mean localization errors under varying label densities from 0.1\% to 1.0\%.}
    \label{fig:Label_Density_Error}
\end{figure}

To evaluate the system-level accuracy and robustness of SkySense, Fig.~\ref{fig:CDF_Error_1_Percent} depicts the cumulative distribution function (CDF) of 3D localization errors under the 1\% label density setting, while Fig.~\ref{fig:Label_Density_Error} illustrates the error trends across a gradient of extreme label scarcity. SkySense significantly outperforms both the fully supervised FC-AE-GPR and the semi-supervised SSLUL. While traditional baselines experience precision collapse as labels are reduced, SkySense maintains a graceful degradation curve. Even at 0.1\% density, it achieves decimeter-level accuracy suitable for practical UAV localization.

The CDF evaluation further reveals the superiority of our generative formulation in suppressing tail errors. Unlike deterministic regression models that inevitably yield invalid average coordinates when confronted with multi-path spatial ambiguity, SkySense probabilistically models the conditional spatial distribution to reliably resolve these conflicting mappings. 

This robustness is fundamentally reinforced by our physics-aware self-supervised strategy. Existing semi-supervised methods, such as SSLUL, rely merely on generic CSI reconstruction, which inherently fails to exploit the continuous spatial-temporal characteristics of the flight trajectory and easily collapses under extreme label scarcity. In contrast, SkySense explicitly leverages the kinematic continuity of UAV flight to extract structured physical priors. This pre-trained knowledge provides a highly robust initialization, guiding the generative backbone to converge accurately with minimal annotated data and significantly alleviating the prohibitive annotation overhead.

\subsection{Ablation Study}
\label{subsec:ablation}
To systematically isolate and validate the performance contributions of individual components, we conduct a comprehensive ablation study, as summarized in Table \ref{tab:combined_results}. All evaluated models follow a unified two-part topology consisting of a shared MLP-based spatial encoder $E_\theta$ and a task-specific backbone $F_\phi$. The framework incorporates three $F_\phi$ architectures—MLP, CNN, and U-Net—to analyze localization efficacy across different representation capacities. Furthermore, a standard DDPM is implemented to isolate the impact of the consistency generative mechanism. The experimental matrix covers fully-supervised and semi-supervised paradigms, incorporating individual evaluations for each base station alongside multi-BS configurations, with all variants constrained to around 1.9M parameters for fairness.
\begin{table*}[htbp]
\centering
\caption{Comparison of 3D Localization Errors Across Different Neural Backbones and Supervision Levels.}
\label{tab:combined_results}
\footnotesize 

\setlength{\tabcolsep}{10pt}
\begin{tabular}{llcccccccc}
\toprule
\multirow{2}{*}{\textbf{Backbone}} & \multirow{2}{*}{\textbf{Method}} & \multicolumn{4}{c}{\textbf{Fully Supervised}} & \multicolumn{4}{c}{\textbf{Semi-Supervised}} \\
\cmidrule(lr){3-6} \cmidrule(lr){7-10}
& & \textbf{BS 1} & \textbf{BS 2} & \textbf{BS 3} & \textbf{3-BS Fusion} & \textbf{BS 1} & \textbf{BS 2} & \textbf{BS 3} & \textbf{3-BS Fusion} \\
\midrule
\multirow{3}{*}{MLP}  
& Discriminative & 174.50 & 178.20 & 184.82 & 150.21 & 74.80  & 79.50  & 85.05  & 65.32 \\
& DDPM          & 302.10 & 278.50 & 322.07 & 282.45 & 315.20 & 342.80 & 335.35 & 290.15 \\
& SkySense      & 65.30  & 67.80  & 64.96  & 56.11  & 28.50  & 26.80  & 35.76  & 13.28 \\
\midrule
\multirow{3}{*}{CNN}  
& Discriminative & 234.80 & 236.50 & 265.12 & 212.56 & 178.50 & 181.20 & 179.91 & 155.40 \\
& DDPM          & 333.50 & 346.10 & 322.46 & 315.30 & 304.80 & 317.50 & 324.86 & 281.55 \\
& SkySense      & 94.20  & 99.50  & 104.39  & 82.34  & 39.80  & 48.50  & 42.86  & 16.93 \\
\midrule
\multirow{3}{*}{U-Net} 
& Discriminative & 98.50  & 111.20 & 101.50  & 85.67  & 38.50  & 46.20  & 48.78  & 30.45 \\
& DDPM          & 282.80 & 286.10 & 292.95 & 264.12 & 292.50 & 276.10 & 263.04 & 245.90 \\
& \textbf{SkySense}   & \textbf{44.50} & \textbf{46.20} & \textbf{44.60} & \textbf{33.75} & \textbf{20.15} & \textbf{26.50} & \textbf{29.36} & \textbf{9.77} \\
\bottomrule
\end{tabular}
\end{table*}

First, evaluating the neural backbone capacity reveals that under the semi-supervised single-BS configuration, the U-Net variant achieves a 20.67 cm error, significantly outperforming MLP (29.02 cm) and CNN (40.72 cm). This gap demonstrates the necessity of the hierarchical encoder-decoder structure in preserving spatial features for accurate coordinate mapping.

Second, we assess the role of the learning paradigm. Transitioning from fully supervised to semi-supervised training reduces the U-Net variant's single-BS error by over 50\%, dropping from 45.10 cm to 20.67 cm. This improvement proves that extracting structural priors from massive unlabeled sequences effectively bridges the representation gap caused by extreme label scarcity.

Beyond structural representation, the results substantiate the critical role of our consistency generative formulation. Comparing SkySense against a standard DDPM baseline under data-scarce conditions, DDPM suffers a catastrophic collapse (245.90 cm). This confirms that step-wise Markov noise estimation struggles to establish robust mappings with limited physical anchors. In contrast, SkySense’s consistency objective enforces a direct trajectory mapping, enabling robust generative modeling without representation collapse.

Finally, we validate the spatial consensus mechanism. Expanding from a single-BS to a 3-BS setup iteratively aggregates local coordinate proposals from distinct geometric viewpoints. This cooperative mechanism effectively eliminates spurious predictions and suppresses multi-path ambiguity, driving the system-level localization error down from 20.67 cm to an unprecedented 9.77 cm.

\subsection{Inference Efficiency}
\label{subsec:inference_speed}
To evaluate practical inference efficiency, we investigate the latency-precision trade-off between the standard step-wise DDPM and our consistency-based framework under the single-BS setting. While precision is paramount, satisfying low latency constraints is equally critical for real-time ISAC deployments. Traditional generative models like DDPM require a long Markov chain of hundreds of iterative denoising steps, rendering them computationally prohibitive for highly dynamic UAV localization.

\begin{figure}[htbp]
    \centering
    \includegraphics[width=0.8\linewidth]{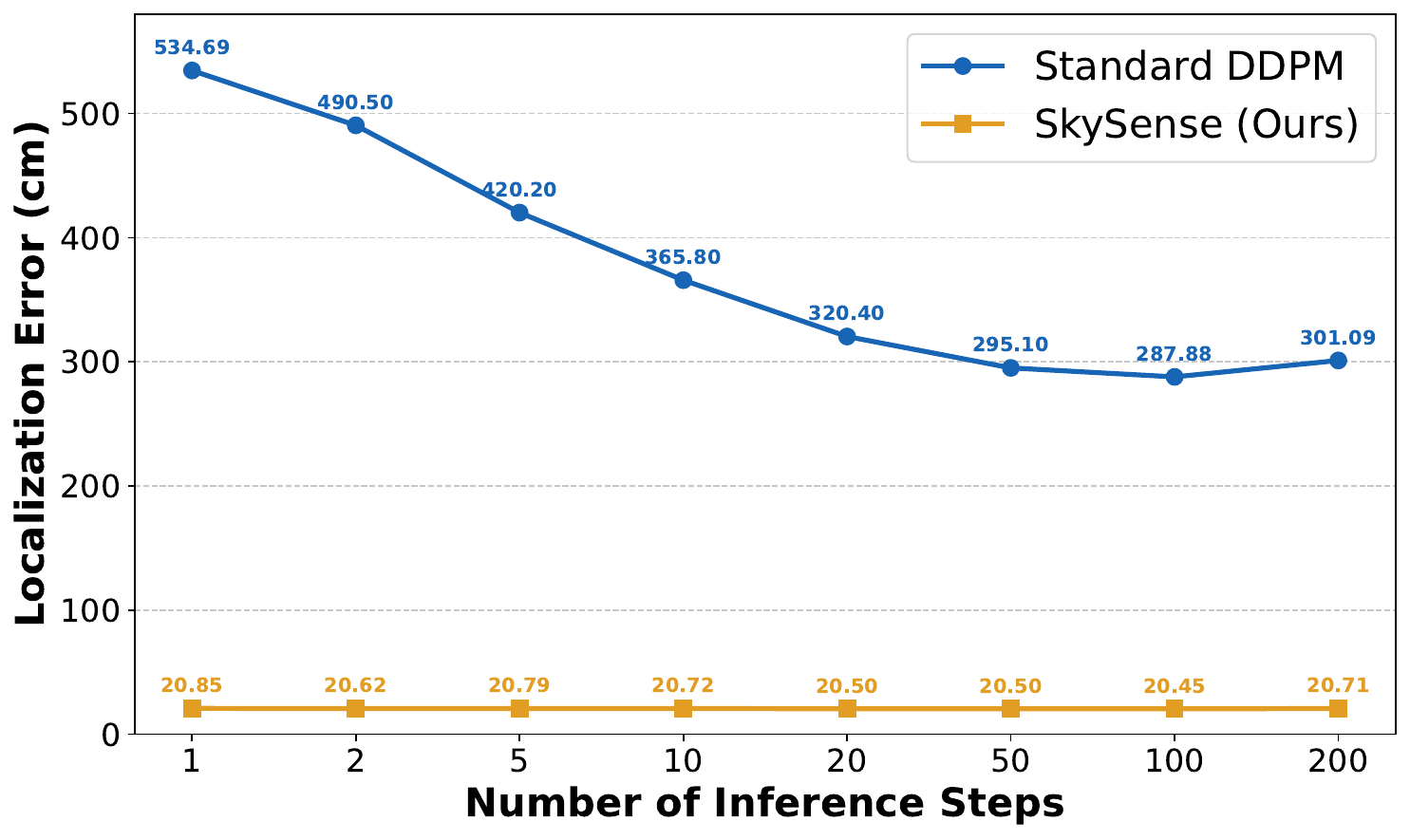}
    \caption{The impact of inference steps on localization error in the single-BS semi-supervised setting.}
    \label{fig:inference_steps}
\end{figure}

As illustrated in Fig.~\ref{fig:inference_steps}, drastically truncating the DDPM trajectory causes exponential error growth, leading the model to ultimately fail at extremely low-step regimes. In sharp contrast, SkySense is explicitly optimized to map any arbitrary noisy state directly to the true physical coordinates, elegantly bypassing step-wise error accumulation. Consequently, SkySense maintains robust accuracy even when the generation process is aggressively compressed to merely 1 to 2 steps. By eliminating the lengthy iterative Markov chain while preserving state-of-the-art precision, SkySense proves to be a highly efficient and scalable solution for real-time UAV localization.

\section{Conclusion}
\label{sec:conclusion}

In this paper, we proposed SkySense, a semi-supervised generative framework tailored for real-time, centimeter-level UAV localization in 6G ISAC networks. By integrating self-supervised spatial representations with a conditional consistency generative backbone, the framework effectively overcomes the precision collapse caused by extreme label scarcity. Furthermore, coupled with a distributed spatial consensus strategy to suppress multipath ambiguity, our architecture requires merely 1 to 2 inference steps to achieve an unprecedented system-level fusion accuracy of 9.77 cm, significantly outperforming existing fully supervised and semi-supervised discriminative baselines. In the future, we aim to leverage high-fidelity ray-tracing simulations to construct large-scale synthetic datasets and employ Sim-to-Real transfer techniques to completely eliminate the reliance on empirical measurements, ultimately enabling a scalable, zero-survey deployment strategy for ubiquitous UAV localization.

\bibliographystyle{IEEEtran}
\bibliography{refs}

\end{document}